\documentstyle[preprint,aps]{revtex} 
\tightenlines
\begin{document} 
\draft      
\preprint{\vbox{\hbox{UCB-PTH-99/56}}, \vbox{\hbox{LBNL-45011}}, 
\vbox{\hbox{hep-ph/0001304}}}
\title{CP violating $Z t \bar{t}$ and $\gamma t \bar{t}$ Couplings at
a Future $e^+ e^-$ Collider} 
\author{S.\ M.\ Lietti$^1$ and Hitoshi Murayama$^{1,2}$\thanks{
  This work was supported in part by the Director, Office of
  Science, Office of High Energy and Nuclear Physics, Division of High
  Energy Physics of the U.S. Department of Energy under Contract  
  DE-AC03-76SF00098 and in part by the National Science Foundation
  under grant PHY-95-14797.  SML was also supported by Funda\c{c}\~ao
  de Amparo \`a Pesquisa do Estado de S\~ao Paulo (FAPESP).
}}

\address{$^1$Theory Group, Lawrence Berkeley National Laboratory\\
         Berkeley, CA 94720, USA.}  
\address{$^2$Department of Physics, University of California\\
  Berkeley, CA 94720, USA.}
\maketitle
\widetext
\begin{abstract}
  The effect of new operators that give rise to CP--violating
  couplings of the type $Z t \bar{t}$ and $\gamma t \bar{t}$ are
  examined at future electron positron Linear Colliders (FLC). The
  impact of these CP-violating interactions over Standard Model
  predictions was studied for the process $e^+ e^- \to t \bar{t}$ with
  the subsequent decays $t \to b l^+ \nu_l$ and $\bar{t} \to \bar{b}
  l^- \bar{\nu}_l$, called as dilepton mode, and $t \to b l^+ \nu_l$
  and $\bar{t} \to \bar{b} \bar{q} q'$ or $t \to b q \bar{q}'$ and
  $\bar{t} \to \bar{b} l^- \bar{\nu}_l$, called as single lepton mode,
  where the final leptons are $l^\pm = e^\pm$ or $\mu^\pm$, and the
  final quarks are $q(q')= u(d)$ or $c(s)$.  Polarized electron beam
  and CP observables and asymmetries are used to impose bounds on the
  anomalous couplings.
\end{abstract}
\pacs{}
\newpage 
\section{Introduction}
\label{int} 
Top quark is the heaviest elementary particle observed to date and 
is hence most sensitive to the mechanism of electroweak symmetry breaking.
The top quark couplings to gauge bosons probe the nature of electroweak 
symmetry breaking and other not well understood aspects of
the electroweak interactions \cite{topgaugecouplings}.
In particular, it would be interesting to investigate whether top
couplings conserve CP, a symmetry so far known to be violated only in
$K$-meson system. Possible CP violating couplings of fermions are
electric dipole type interactions with the electromagnetic field and
the analogous ``weak'' dipole coupling to the $Z$ field. 
These can arise, for instance, in certain models of CP
violation like the two-Higgs-doublet model \cite{exthiggssec},
in the minimal supersymmetric standard model (MSSM) at one-loop level 
\cite{mssm} or in its next-to minimal extension (NMSSM) at tree level
\cite{nmssm} even though the order of magnitudes of their estimate is
probably well below the experimental sensitivity.

We, however, find that the supersymmetric contribution to the
CP-violating top couplings can be sizable.  For instance, a gluino
exchange together with the stop left-right mixing would produce the
electric dipole moment of the order of $e(\alpha_{s}/\pi) {\rm
  Im}(A^{*} M_{3}) m_{t}/m_{\tilde{t}}^{4}$ and the form factors
defined in Section \ref{formfactor} can easily be of a few percents if
$m_{\tilde{t}} \sim m_{t}$ which is still allowed.  Note that the
constraints from the neutron and electron electric dipole moments do
not restrict the trilinear coupling $A$ for the stop unless specific
assumptions such as the universal trilinear coupling is made.

In this paper, we do not restrict ourselves to any particular model, 
but parametrize the CP violation in terms of convenient effective form 
factors proportional to the electric and weak dipole moments of the 
top-quark.
 
A high-energy future linear $e^+ e^-$ collider (FLC) will provide a
very impressive tool to investigate the properties of the top-quark. 
Since the mass of the top-quark is very high ($m_t=174.3\pm5.1$ GeV) 
\cite{pdgmtop}, its weak decay takes place before it can hadronize and 
hence it can be studied in a much cleaner way than other quarks. Moreover, 
since all theories involving CP violation effects in the electroweak 
coupling of fermions are expected to be proportional to their mass, 
the top-quark is a privileged candidate for observing such effects 
\cite{FC1}.

In this paper we will study possible CP violating effects due to anomalous
form factors to the vertex $(Z,\gamma)t \bar{t}$ \cite{GLK-GAL-CPY,GAL-CPY} 
in the top-quark production at an $e^+ e^-$ collider, {\it i.e.}\/, 
$e^+ e^- \to Z,\gamma \to t \bar{t}$.
These form factors are presented in Section \ref{formfactor}.

There have been several studies to measure possible CP violating effects 
due to non standard $Zt\bar{t}$ and $\gamma t\bar{t}$ couplings. Various 
experiments have been suggested to perform these measurements by making use 
of CP-odd quantities (see Ref. \cite{FC1,FC2} and references therein). In this
paper we study the impact of CP violating $Zt\bar{t}$ and 
$\gamma t\bar{t}$ couplings using two sets of CP-odd observables
\cite{FC1,FC2,WB-AB-MF}, by 
studying their expectation values and their corresponding asymmetries
defined in Section \ref{obs_asy}.


Moreover, effects of a possible highly polarized electron beam ($\pm$90\%) 
at FLC will be considered in our analyses of CP violating $Zt\bar{t}$ 
and  $\gamma t\bar{t}$ couplings. Our results are presented in 
Section \ref{results}. Finally, we draw our conclusions in Section 
\ref{conclusions}.
  
\section{The General Form Factors} 
\label{formfactor} 
In order to study the effects of CP violating form factors to the 
vertex $(Z,\gamma)t \bar{t}$, we use the most general form 
factors for the coupling of $t$ and $\bar{t}$ 
with either $Z$ or $\gamma$ defined in Ref.\cite{GLK-GAL-CPY}, 
\begin{eqnarray}
\Gamma^\mu_{V t \bar{t}}=ig&&\left[\gamma^\mu (F_1^{V(L)}P_- + F_1^{V(R)}P_+)
-\frac{i \sigma^{\mu \nu} k_\nu}{m_t}( F_2^{V(L)}P_- + F_2^{V(R)}P_+)\right.
\nonumber \\
&&\left. + k^\mu( F_3^{V(L)}P_- + F_3^{V(R)}P_+)\right],
\label{eq1}
\end{eqnarray}
were $P_{\pm}=\frac{1}{2}\left(1 \pm \gamma_5 \right)$, $i\sigma^{\mu \nu} =
- \frac{1}{2} \left[ \gamma^\mu,\gamma^\nu \right]$, $m_t$ is the top mass,
 $k^\mu$ is the momentum of the gauge boson $V$ and is taken by convention 
to be directed into the vertex. $V$ can be the $Z$ gauge boson or photon $A$, 
and the $F$'s are the form factors for $V$. When $V=A$, $F_3^{A(L)}$ and 
$F_3^{A(R)}$ have to vanish as a result of gauge invariance (or current 
conservation). For a $Z$ boson which is on shell or coupled to massless 
fermions, the $F_3^{Z(L)}$ and $F_3^{Z(R)}$ contributions vanish. In our 
case we will ignore these $F_3$ contributions. The Standard Model values for
the form factors at tree level are:
\begin{eqnarray}
&&F_{1_{SM}}^{Z(L)}= \frac{1}{\cos \theta_W}\left[\frac{1}{2}-
\frac{2}{3}\sin^2\theta_W \right]
\;\;,\;\;
F_{1_{SM}}^{Z(R)}=\frac{1}{\cos \theta_W}\left[-
\frac{2}{3}\sin^2\theta_W \right]\;\;,
\nonumber \\  
&&F_{1_{SM}}^{A(L)}=F_{1_{SM}}^{A(R)}=\frac{2}{3}\sin\theta_W\;\;,
\label{eq2} \\  
&&F_{2_{SM}}^{Z(L)}=F_{2_{SM}}^{Z(R)}=F_{2_{SM}}^{A(L)}=
F_{2_{SM}}^{A(R)}=0\;\;,
\nonumber 
\end{eqnarray}
where $\theta_W$ is the weak mixing angle. 

Applying the Gordon decomposition, equation (\ref{eq1}) becomes
\begin{eqnarray}
\Gamma^\mu_{V t \bar{t}}=\frac{ig}{2} \left[\gamma^\mu (A^V-B^V\gamma^5)
+\frac{t^\mu-\bar{t}^\mu}{2} (C^V-D^V\gamma^5)\right],
\label{eq3}
\end{eqnarray}
where
\begin{eqnarray}
A^V&=&F_1^{V(L)}+F_1^{V(R)} -2(F_2^{V(L)}+F_2^{V(R)}), 
\nonumber \\
B^V&=&F_1^{V(L)}-F_1^{V(R)}, 
\nonumber \\
C^V&=& \frac{2}{m_t} \left(F_2^{V(L)}+F_2^{V(R)} \right), 
\label{eq4} \\ 
D^V&=& \frac{2}{m_t} \left(F_2^{V(L)}-F_2^{V(R)} \right).
\nonumber
\end{eqnarray}
In equation (\ref{eq3}), $t^\mu$ ($\bar{t}^\mu$) is the momentum of the 
outgoing $t$ ($\bar{t}$). The Standard Model values at tree level of these
last set of form factors are
 \begin{eqnarray}
A^Z_{SM}&=&\frac{1}{\cos \theta_W}\left[\frac{1}{2}-\frac{4}{3}\sin^2 
\theta_W \right] \;\;,\;\; 
A^A_{SM}=\frac{4}{3}\sin \theta_W\;\;, 
\nonumber \\
B^Z_{SM}&=&\frac{1}{2 \cos \theta_W}\;\;,\;\;
B^A_{SM}=0\;\;,
\\
C^Z_{SM}&=&C^A_{SM}=D^Z_{SM}=D^A_{SM}=0.
\nonumber
\end{eqnarray} 
Beyond the tree level, all of them except $D^V$ ($V=Z$ or $A$),
which controls the CP violation, have contributions due to loop 
corrections in the SM provided we ignore the small CP-violating 
effects which reside in the Yukawa couplings that govern  the interactions
between the Higgs boson and the quarks \cite{GAL-CPY}. These CP-violating
amplitudes in the Cabibbo-Kobayashi-Maskawa model are typically 
suppressed by a factor of order $10^{-12}$ \cite{CJ}.

Since we are interested in possible non-standard CP-violating effects of the 
vertex $(Z,\gamma)t\hat{t}$, we will analyze the impact of the form factor
$D^V$ in the process $e^+ e^- \to (Z,\gamma) \to t \bar{t}$. For convenience, 
we define a dimensionless CP-violating coupling
constants $d^V=(m_t/2) D^V$ and from equation (\ref{eq4}) it is 
obvious that
\begin{eqnarray}
d^V = (F_2^{V(L)}-F_2^{V(R)}) . 
\label{eq7}
\end{eqnarray} 
The Standard Model value at tree level for $d^V$ is zero. The impact of
non-vanishing values for $d^V$ in the processes (\ref{eq6}) will be
studied here 
through the analysis of the processes
\begin{eqnarray} 
e^+ e^- &\to& t(\to b l^+ \nu_l)\;\; 
\bar{t}(\to \bar{b} l^-  \bar{\nu}_l)\;\;,
\label{eq6} 
\end{eqnarray} 
where the final leptons $l^\pm$ are $e^\pm$ or $\mu^\pm$, 
called dilepton mode and, 
\begin{eqnarray} 
e^+ e^- &\to& t(\to b q \bar{q}')\;\; 
\bar{t}(\to \bar{b} l^-  \bar{\nu}_l)\;\;,
\label{eq6.1} \\
e^+ e^- &\to& t(\to b l^+ \nu_l)\;\; 
\bar{t}(\to \bar{b} \bar{q} q')\;\;,
\label{eq6.2}
\end{eqnarray}  
where the final quarks $q(q')$ are the up(down)-quarks 
$u(d)$ or $c(s)$, called single lepton mode. Eq. (\ref{eq6.1}) will
be called sample ${\cal T}$, while Eq. (\ref{eq6.2}) will be called
sample $\bar{\cal T}$ of the single lepton decay mode.

In order to compute these contributions, we have
incorporated  all anomalous couplings in HELAS--type \cite{helas}
Fortran subroutines. These new subroutines were used to adapt a
Madgraph \cite{madgraph} output to include all the anomalous
contributions. We have checked that our code
is able to reproduce the results for
helicity amplitudes Eq.~(2.8) of Ref.~\cite{GAL-CPY}.
We employed Vegas \cite{vegas}  to perform the Monte Carlo 
phase space integration with the appropriate cuts to obtain 
the differential and total cross sections of the processes (\ref{eq6}),
(\ref{eq6.1}), and (\ref{eq6.2}). 

\section{Observables and Asymmetries}
\label{obs_asy} 

The effects of CP-violating form factors of the vertex $(Z,\gamma)t\bar{t}$ 
can be traced through the analysis of the behavior
of some convenient CP observables. For the dilepton decay channel
of the top-quark pair production at FLC we will consider two sets of 
observables. 

The first set of observables was defined in Ref.\cite{FC1,FC2} in order to
study the impact of CP-invariant form factor of the vertex
$Vt\bar{t}$ ($V=Z,\gamma$). It consists in the following two observables:
\begin{eqnarray}
O_1 &=& (\hat{p}_b \times \hat{p}_{\bar{b}}) \cdot \hat{p}_{e^+} \;\;,
\label{obs1} \\
O_2 &=& (\hat{p}_b + \hat{p}_{\bar{b}}) \cdot \hat{p}_{e^+} \;\;,
\label{obs2}
\end{eqnarray}
where $\hat{p}_b$ and $\hat{p}_{\bar{b}}$ are the $b$, $\bar{b}$ momentum 
directions in the $e^+ e^-$ CM frame and $\hat{p}_{e^+}$ is the
momentum direction of the positron.
The observable $O_1$ is CP odd but CPT even, and probes the imaginary 
part of the CP-violating form factors [{\it Im}($d^{Z,\gamma}$)], while 
the observable $O_2$ is both CP and CPT odd and  probes the real part of 
the CP-violating form factors [{\it Re}($d^{Z,\gamma}$)]. A  CPT-odd
observable can only have a non-zero value in the presence of an 
absorptive part of the amplitude \cite{cpteven}.\footnote{Here and 
below, we have the ``naive T'' in mind where spins and momenta are 
reversed but the initial and final states are not interchanged.  
Therefore, CPT-odd observables do not imply the true CPT violation 
which is of course impossible in quantum field theories.}

The second set was defined in Ref.\cite{WB-AB-MF} in order to study
effects of Higgs sector CP violation in top-quark pair production.
It consists in the following two observables:
\begin{eqnarray}
Q_1 &=& \hat{p}_t \cdot \hat{q}_+ - \hat{p}_{\bar{t}} \cdot \hat{q}_- \;\;,
\label{obsq1} \\
Q_2 &=& \frac{1}{2} (\hat{p}_t - \hat{p}_{\bar{t}}) \cdot
(\hat{q}_- \times \hat{q}_+) \;\;,
\label{obsq2}
\end{eqnarray}
where $\hat{p}_t$ and $\hat{p}_{\bar{t}}$ are the $t$, $\bar{t}$ momentum 
directions in the $e^+ e^-$ CM frame and $\hat{q}_-$ and $\hat{q}_+$ are
the $l^+$, $l^-$ momentum directions in the  $t$ and $\bar{t}$ rest frames,
respectively. The observable $Q_1$ is CP odd but T even, i. e. do not
change sign under a naive T transformation, and probes the real 
part of the CP-violating form factors [{\it Re}($d^{Z,\gamma}$)], while 
the observable $O_2$ is both CP and T odd and probes the imaginary part of 
the CP-violating form factors [{\it Im}($d^{Z,\gamma}$)]. 

For the single lepton decay channel of the top-quark pair production at 
FLC we will also consider two sets of observables.  The first set of 
observables consists of the observables of Eqs. (\ref{obs1}) and 
(\ref{obs2}), {\it i.e.}\/, the same set of observables for the dilepton decay 
channel.

However, the second set of observables must not be the same of the dilepton 
decay channel [Eqs. (\ref{obsq1}) and (\ref{obsq2})]. Instead we use 
the following observables:

For the sample 
$e^+e^- \to t (\to b l^+ \nu_l)  \bar{t}(\to \bar{b} q \bar{q}')$
(sample ${\cal T}$) they define
\begin{eqnarray}
Q_1^{(t)} &=& \hat{p}_t \cdot \hat{q}_+  \;\;,
\label{obsq1t} \\
Q_2^{(t)} &=& \hat{p}_t \cdot (\hat{q}_+ \times \hat{q}_{\bar{b}})\;\;,
\label{obsq2t}
\end{eqnarray}
where $\hat{q}_{\bar{b}}$ is the momentum direction of the $\bar{b}$ quark 
jet in the $\bar{t}$ quark rest frame, while for the process 
$e^+e^- \to t (\to b \bar{q} q')  \bar{t}(\to \bar{b} l^- \bar{\nu}_l)$
(sample $\bar{\cal T}$),
\begin{eqnarray}
Q_1^{(\bar{t})} &=& \hat{p}_{\bar{t}} \cdot \hat{q}_-  \;\;,
\label{obsq1tb} \\
Q_2^{(\bar{t})} &=& \hat{p}_{\bar{t}} \cdot (\hat{q}_- \times \hat{q}_b)\;\;,
\label{obsq2tb}
\end{eqnarray}
where $\hat{q}_b$ is the momentum direction of the $b$ quark jet in the 
$t$ quark rest frame. Taking both samples one can define the quantities
\begin{eqnarray}
\epsilon_1 = <Q_1^{(t)}> - <Q_1^{(\bar{t})}> \;\;,
\label{epsilon1} \\
\epsilon_2 = <Q_2^{(t)}> + <Q_2^{(\bar{t})}> \;\;.
\label{epsilon2}
\end{eqnarray}
The quantity $\epsilon_1$ probes the real part of the CP-violating form 
factors [{\it Re}($d^{Z,\gamma}$)], while $\epsilon_2$ probes the imaginary 
part of the CP-violating form factors [{\it Im}($d^{Z,\gamma}$)].

We also define corresponding asymmetries which should be experimentally
more robust than equations (\ref{obs1},\ref{obs2},\ref{obsq1},\ref{obsq2},
\ref{epsilon1},\ref{epsilon2}), because only the 
signs of $O_{1,2}$, $Q_{1,2}$, and $\epsilon_{1,2}$ have to be measured.
For the first set of  observables, we define the asymmetry 
for both single and di- lepton decay channels, as follows
\begin{eqnarray}
A_{O_{1,2}}=\frac{N(O_{1,2}>0)-N(O_{1,2}<0)}
{N(O_{1,2}>0)+N(O_{1,2}<0)}\;\;, 
\label{asymmetry}
\end{eqnarray}
where $N$ is the number of $t \bar{t}$ events in the
single and di- lepton decay channels.

For the second set of  observables we define the asymmetry as follows:
for the dilepton decay channel,
\begin{eqnarray}
A_{Q_{1,2}}=\frac{N(Q_{1,2}>0)-N(Q_{1,2}<0)}
{N(Q_{1,2}>0)+N(Q_{1,2}<0)}\;\;, 
\label{asymmetry_Q}
\end{eqnarray}
where $N$ is the number of $t \bar{t}$ events in the
dilepton decay channels.

For the single lepton decay channel,
\begin{eqnarray}
A(\epsilon_1) = \frac{N_{\cal T}(Q_1^{(t)}>0)-N_{\cal T}(Q_1^{(t)}<0)}
{N_{\cal T}} - 
\frac{N_{\bar{\cal T}}(Q_1^{(\bar{t})}>0)-N_{\bar{\cal T}}(Q_1^{(\bar{t})}<0)}
{N_{\bar{\cal T}}} \;\;,
\label{asymm_epsilon1} \\
A(\epsilon_2) = \frac{N_{\cal T}(Q_2^{(t)}>0)-N_{\cal T}(Q_2^{(t)}<0)}
{N_{\cal T}} + 
\frac{N_{\bar{\cal T}}(Q_2^{(\bar{t})}>0)-N_{\bar{\cal T}}(Q_2^{(\bar{t})}<0)}
{N_{\bar{\cal T}}} \;\;,
\label{asymm_epsilon2}
\end{eqnarray}
where $N_{\cal T}$ and $N_{\bar{\cal T}}$ are the number of $t\bar{t}$ events 
in samples ${\cal T}$ and $\bar{\cal T}$, respectively. 

The sensibility of non-null values of the CP-violating form factors 
$d^{Z,\gamma}$ over these two sets of observables and correspondent 
asymmetries are summarized in TABLE \ref{table1}.

\section{Results}
\label{results}
The impact of the CP-violating form factors described in Section 
\ref{formfactor} in the top-quark pair production and subsequent
decay into 2 jets plus 2 leptons (dilepton mode), and into 
4 jets plus 1 lepton (single lepton mode) is analyzed for a  FLC with CM
energy of 500 GeV. 
Polarization effects of the electron beam is also considered. We assume two 
runs at FLC, one with $90\%$ left hand polarized electrons 
(${\cal P}^-_{e^-}$) and the other run with  $90\%$ right hand polarized 
electrons (${\cal P}^+_{e^-}$), both with integrated luminosity of 
50 fb$^{-1}$. We have considered $m_t=$ 175 GeV in our analysis.

A discussion concerning event selection and backgrounds, that can be found 
in Ref.\cite{background} and references therein, is briefly summarized here.
The $t\bar{t}$ cross section at an FLC with $\sqrt{s}=500$ GeV is roughly 
0.5 pb. On the other hand, the cross section for lepton and light quark pairs 
is about 16 pb, while for $W^+W^-$ production is about 8 pb. The emphasis 
of most event selection strategies has been to take advantage of the 
multi-jet topology of the roughly 90\% of  $t\bar{t}$ events with 4 or 6 jets 
in the final state. Therefore, cuts on thrust or number of jets 
drastically reduces the light fermion pair background. In addition, one can 
use the multi-jet mass constraints $M(\mbox{jet-jet}) \approx M_W$ and
$M(\mbox{3-jet}) \approx m_t$ for the cases involving $t\to b q q'$.
The background due to $W$-pair production is the most difficult to eliminate. 
However, in the limit that the electron is fully right-handed polarized, 
the $W^+W^-$ cross section is reduced to about 30 fb. Hence, even though the 
beam polarization will not reach 100\%, this allows for experimental control 
and measurement of the background. Another important technique that can be 
used is that of precision vertex detection. The small and stable interaction
point of  linear $e^+e^-$ colliders, along with the small beam sizes and
bunch-structure timing, make then ideal for pushing the techniques of 
vertex detection.

The Standard Model total cross sections of $t\bar{t}$ production at FLC with 
$\sqrt{s}=$ 500 GeV obtained by our Monte Carlo simulation are:
\begin{eqnarray}
\sigma_{e^+ e^-\to t\bar{t}}({\cal P}^-_{e^-}) &=&
777.3(3)\;\;\mbox{fb} \;\;  ,
\\
\sigma_{e^+ e^-\to t\bar{t}}({\cal P}^+_{e^-})&=&
373.8(1)\;\;\mbox{fb} \;\; .
\end{eqnarray}

We conservatively assume [$W^-(\to l^- \bar{\nu}_l) W^+(\to l^+ \nu_l)$],
[$W^-(\to l^- \bar{\nu}_l) W^+(\to q \bar{q}')$],
[$W^-(\to \bar{q} q') W^+(\to l^+ \nu_l)$] tagging efficiencies
of about $80\%$ and a $b$ and $\bar{b}$ tagging efficiency also of $80\%$.  
The overall $b-$, $\bar{b}-$, and $W-$ tagging efficiency would then
be about $(80\%)^3 = 51.2\%$. In our calculations we consider an overall 
tagging efficiency of $50\%$ ($f_{eff}=0.5$). 
Considering the leptons being only electron and muon, the branching ratio 
of the dilepton decay mode is $BR=\frac{4}{81}$. For the single lepton decay,
when the final quarks are the quarks up, down, charm, and strange, 
the branching ratio  is $BR=\frac{24}{81}$.
The number of events in each decay mode of the top-quark pair production 
at FLC is given by $N=\sigma.{\cal L}.f_{eff}.BR$ and is shown in TABLE 
\ref{table2}.  

\subsection{Expectation Values}
The observables defined in Section \ref{obs_asy} 
acquire non-vanishing expectation values  
in the presence of CP-violating anomalous couplings $Z t \bar{t}$ and 
$\gamma t \bar{t}$. Expectation values of observables are defined as usual by
\begin{eqnarray}
\left\langle {\cal O} \right\rangle = 
\frac{\int d\sigma {\cal O}}{\int d\sigma} \;\;.
\label{expcval}
\end{eqnarray}

To be statistically significant, the expectation values of an observable
${\cal O}$  must be larger then its expected natural variances 
$\langle({\cal O}-\langle {\cal O}\rangle)^2\rangle$.\footnote{Of
  course $\langle {\cal O} \rangle =0$ in the Standard Model.}
A signal of $\eta$ standard deviations is obtained for a sample of $N$ events 
if
\begin{eqnarray}
\left\langle {\cal O} \right\rangle \geq 
\eta \sqrt{\frac{\left\langle {\cal O}^2 \right\rangle}{N}} 
\;\; .
\label{significance}
\end{eqnarray}
 
In order to obtain bounds on the anomalous form factors $d^{Z,\gamma}$
we have evaluated numerically, for the first set of observables, the fraction
\begin{eqnarray}
{\cal F}_{O_{1,2}}= \frac{\left\langle O_{1,2} \right\rangle}
{\sqrt{\left\langle O_{1,2}^2 \right\rangle}} \;\;,
\label{fraction_O}
\end{eqnarray}
for different values of the form factors $d^{Z,\gamma}$ for both
dilepton and single lepton modes.

For the second set of observables we have evaluated numerically, for
the dilepton mode, the fractions,
\begin{eqnarray}
{\cal F}_{Q_{1,2}}= \frac{\left\langle Q_{1,2} \right\rangle}
{\sqrt{\left\langle Q_{1,2}^2 \right\rangle}} \;\;,
\label{fraction_Q}
\end{eqnarray}
while for the single lepton mode we have evaluated numerically the fractions,

\begin{eqnarray}
{\cal F}_{Q_{1,2}^{(t)}}= 
\frac{\left\langle Q_{1,2}^{(t)} \right\rangle}
{\sqrt{\left\langle {Q_{1,2}^{(t)}}^2 \right\rangle}}
\; \; \; , \; \; \;
{\cal F}_{Q_{1,2}^{(\bar{t})}}= 
\frac{\left\langle Q_{1,2}^{(\bar{t})} \right\rangle}
{\sqrt{\left\langle {Q_{1,2}^{(\bar{t})}}^2 \right\rangle}},
\nonumber
\end{eqnarray}
respectively for the samples ${\cal T}$ and $\bar{\cal T}$ 
for different values of the form factors $d^{Z,\gamma}$. Then
we evaluate the following quantities,
\begin{eqnarray}
{\cal F}_{\epsilon_{1}} = {\cal F}_{Q_{1}^{(t)}} - 
{\cal F}_{Q_{1}^{(\bar{t})}} \;\;,
\label{fraction_epsilon1} \\
{\cal F}_{\epsilon_{2}} = {\cal F}_{Q_{2}^{(t)}} +
{\cal F}_{Q_{2}^{(\bar{t})}} \;\;.
\label{fraction_epsilon2}
\end{eqnarray}

A 95\% CL bound is obtained when $\eta=\pm1.96$, so calling
by ${\cal F}$ the quantities of Eqs. (\ref{fraction_O}, \ref{fraction_Q}, 
\ref{fraction_epsilon1}, \ref{fraction_epsilon2}), we have to observe
\begin{eqnarray}
|{\cal F}|
\geq \frac{|\eta|}{\sqrt{N_{events}}}=\frac{1.96}{\sqrt{N_{events}}},
\end{eqnarray}
where the total number of events of both samples, for each polarization 
mode of the electron beam, which is presented in TABLE \ref{table2}.
Our results are presented in TABLE \ref{table4} for the dilepton decay 
mode and in TABLE \ref{table6} for the single lepton decay mode.

\subsection{Asymmetries}

The asymmetry in the observable ${\cal O}$ is defined by
\begin{equation}
  A_{\cal O} \equiv \frac{N({\cal O}>0)-N({\cal O}<0)}
  {N({\cal O}>0)+N({\cal O}<0)} \, .
\end{equation}
The asymmetry is predicted to be zero in the Standard Model for all
observables defined in Section \ref{obs_asy}.  The Gaussian
fluctuation in the asymmetry is given by
\begin{equation}
  \langle (A_{\cal O}-\langle A_{\cal O}\rangle)^2 \rangle
  = 4 \frac{N({\cal O}>0)N({\cal O}<0)}{(N({\cal O}>0)+N({\cal
  O}<0))^{3}} = \frac{1}{N_{events}},
\label{asy-eve}
\end{equation}
where vanishing asymmetry $N({\cal O}>0)=N({\cal O}<0)$ was assumed in
the last equality.  

Hence, from Eqs. (\ref{asymmetry},\ref{asymmetry_Q},\ref{asymm_epsilon1},
\ref{asymm_epsilon2}), and (\ref{asy-eve}), a 95\% CL 
deviation is obtained when one measures
the asymmetry
\begin{eqnarray}
A_{\cal O}^{95\%CL}=\frac{\pm1.96}{\sqrt{N_{events}}}\;\;.
\label{asy-eve2}
\end{eqnarray}
We present in TABLE \ref{table3} the value of the quantity $A_{\cal O}$
needed to obtain a 95\% CL deviation from the Standard Model prediction
considering the total number of events $N$ presented in TABLE \ref{table2} 
for each decay channel mode of the top-quark pair production at FLC.
Once again, we have evaluated numerically the quantity $A_{\cal O}$ for
different values of the form factors $d^{Z,\gamma}$ in order to
obtain a 95\% CL CP violating signal. Our results are presented in 
TABLE \ref{table5} for the dilepton decay mode and in 
TABLE \ref{table7} for the single lepton decay mode.

\subsection{Improving the Limits}
In order to improve the limits obtained for each polarization mode
of the electron beam, we combine the results of both modes.
We define,
\begin{eqnarray}
{\cal F}^\pm &=& {\cal F}({\cal P}^-_{e^-}) \pm {\cal F}({\cal P}^+_{e^-})\;\;,
\nonumber \\
A_{\cal O}^\pm &=& A_{\cal O}({\cal P}^-_{e^-}) \pm 
A_{\cal O}({\cal P}^+_{e^-})\;\;.
\nonumber 
\end{eqnarray}
The number of events for these new quantities is 
\begin{eqnarray}
N_{events}=N_{events}({\cal P}^-_{e^-})+N_{events}({\cal P}^+_{e^-}).
\nonumber 
\end{eqnarray}
TABLES \ref{table4}, \ref{table5}, \ref{table6}, and \ref{table7} show 
the improved limits for these quantities. 

\section{Conclusions}
\label{conclusions}

The effect of new operators that give rise to CP--violating
couplings of the type $Z t \bar{t}$ and $\gamma t \bar{t}$ were
examined at future electron positron Linear Colliders (FLC). The
impact of these CP-violating interactions over Standard Model
predictions was studied for the process $e^+ e^- \to t \bar{t}$ with
the subsequent decays into a pair of $b$ jets plus four leptons 
(dilepton mode), and decays into a pair of $b$ jets plus a pair of 
light quark jets plus a pair of leptons (single lepton mode).  

Polarized electron beam and two set of CP observables and asymmetries 
were used to impose bounds on the anomalous couplings. The first set of 
observables was defined in Ref.\cite{FC1,FC2}, while the second one 
was defined in Ref.\cite{WB-AB-MF}. 

Our evaluations show that, for the dilepton mode, the second set of 
observables provides  better results than the first one. This is more 
evident for the real part of the anomalous form factors $d^{Z,\gamma}$, 
as one can see in Tables \ref{table4} and \ref{table5}. However, for the 
single lepton mode, the first set of observables is the one that provides 
better results. Once again, this is more evident for the real 
part of the anomalous form factors $d^{Z,\gamma}$, as shown in 
Tables \ref{table6} and \ref{table7}.


According to the statement that the study of the asymmetries is 
experimentally more robust than evaluation of expectation values because
only the signs of the observables have to be measured, the measurement of 
asymmetries can be an important tool in the search for CP-violating effects 
in $t \bar{t}$ production at a future linear $e^+ e^-$ collider.
Our results show that the bounds obtained for the expectation values 
analyses on Tables \ref{table4} and \ref{table6} and  the bounds from the 
asymmetry  analyses on \ref{table5} and \ref{table7} are very similar.
So we conclude that the study of the asymmetries should be experimentally 
easier, with good results.  Furthermore the sensitivity approaches the order 
of magnitudes which can arise in supersymmetric theories.

\acknowledgments
This work was supported in part by the Director, Office of
  Science, Office of High Energy and Nuclear Physics, Division of High
  Energy Physics of the U.S. Department of Energy under Contract  
  DE-AC03-76SF00098 and in part by the National Science Foundation
  under grant PHY-95-14797.  SML was also supported by Funda\c{c}\~ao
  de Amparo \`a Pesquisa do Estado de S\~ao Paulo (FAPESP).

\begin{table}
\begin{tabular}{||c||c||c||}
Form Factor & Dilepton Mode & Single Lepton Mode \\
\hline
\hline
{\it Re}$[d^{Z,\gamma}]$ & $O_2$ and $Q_1$ & $O_2$ and $\epsilon_1$ \\
\hline
{\it Im}$[d^{Z,\gamma}]$ & $O_1$ and $Q_2$ & $O_1$ and $\epsilon_2$
\end{tabular}
\caption{Sensibility of the observables $O_{1,2}$, $Q_{1,2}$ and 
$\epsilon_{1,2}$ to the CP-invariant form  factor $d^{Z,\gamma}$.}
\label{table1}

\end{table} 
\begin{table}
\begin{tabular}{||c||c||c||}
Polarization Mode & Dilepton Mode & Single Lepton Mode \\
\hline
\hline
${\cal P}^-_{e^-}$ &  960 & 5758 \\
\hline
${\cal P}^+_{e^-}$ &  461 & 2769 \\
\hline
${\cal P}^-_{e^-}+{\cal P}^+_{e^-}$ & 1421 & 8527  
\end{tabular}
\caption{Expected number of events per each channel decay mode of
$t \bar{t}$ production at FLC with $\sqrt{s}=500$GeV, ${\cal L}=50$fb$^{-1}$,
and a conservative overall tagging efficiency of 50\%.}
\label{table2}
\end{table}

\begin{table}
\begin{tabular}{||c||c||c||}
Polarization Mode & Dilepton Mode & Single Lepton Mode \\
\hline
\hline
${\cal P}^-_{e^-}$ &  $\pm$6.33 \% &  $\pm$2.58 \%  \\
\hline
${\cal P}^+_{e^-}$ &  $\pm$9.13 \% &  $\pm$3.72 \%   \\
\hline
${\cal P}^-_{e^-}+{\cal P}^+_{e^-}$ &  $\pm$5.20 \%  &  $\pm$2.12 \%  
\end{tabular}
\caption{Expected values for the fraction ${\cal F}$ or for the asymmetry 
$A_{\cal O}$ of a CP-observable for a 95\% CL deviation from the Standard 
Model prediction.}
\label{table3}
\end{table}   

\begin{table}
\begin{tabular}{||c||c||c||c||c||}
Expected Value ${\cal F}$ & 
${\cal P}^-_{e^-}$  & ${\cal P}^+_{e^-}$  & 
${\cal P}^-_{e^-} + {\cal P}^+_{e^-}$ & 
${\cal P}^-_{e^-} - {\cal P}^+_{e^-}$ \\ 
\hline
\hline
{\it Im}($d^\gamma$) from $O_1$ & ($-$0.130 , 0.129) & 
($-$0.181 , 0.178) & ($-$2.59 , 2.54) & ($-$0.053 , 0.052)\\
\hline
{\it Im}($d^\gamma$) from $Q_2$ & ($-$0.119 , 0.121) & 
($-$0.173 , 0.173) & ($-$0.049 , 0.050) & ($-$158 , 160)\\
\hline 
{\it Im}($d^Z$) from $O_1$ & ($-$0.192 , 0.193) & 
($-$0.260 , 0.257) & ($-$0.077 , 0.076) & ($-$2.19 , 2.11)\\
\hline 
{\it Im}($d^Z$) from $Q_2$ & ($-$0.192 , 0.188) & 
($-$0.366 , 0.366) & ($-$0.624 , 0.612) & ($-$0.091 , 0.088)\\
\hline 
\hline 
{\it Re}($d^\gamma$) from $O_2$ & ($-$0.300 , 0.299) & 
($-$0.260 , 0.259) & ($-$0.093 , 0.092) & ($-$0.372 , 0.370)\\
\hline 
{\it Re}($d^\gamma$) from $Q_1$ & ($-$0.127 , 0.128) & 
($-$0.176 , 0.174) & ($-$0.051 , 0.051) & ($-$1.96 , 1.91)\\
\hline 
{\it Re}($d^Z$) from $O_2$ & ($-$0.472 , 0.461) & 
($-$0.516 , 0.517) & ($-$1.25 , 1.29) & ($-$0.169 , 0.164)\\
\hline
{\it Re}($d^Z$) from $Q_1$ & ($-$0.184 , 0.184) & 
($-$0.387 , 0.383) & ($-$0.487 , 0.495) & ($-$0.090 , 0.089)
\end{tabular}
\caption{Expected 95\% CL bounds on $d^{Z,\gamma}$ from the expectation value 
(fraction ${\cal F}$) of the observables at FLC for the dilepton decay mode.}
\label{table4}
\end{table}    

\begin{table}
\begin{tabular}{||c||c||c||c||c||}
Asymmetry &
${\cal P}^-_{e^-}$  & ${\cal P}^+_{e^-}$  & 
${\cal P}^-_{e^-} + {\cal P}^+_{e^-}$ & 
${\cal P}^-_{e^-} - {\cal P}^+_{e^-}$ \\  
\hline
\hline
{\it Im}($d^\gamma$) from $A(O_1)$ & ($-$0.155 , 0.157) & ($-$0.201 , 0.199) & 
($-$1.04 , 1.01) & ($-$0.061 , 0.060)\\
\hline
{\it Im}($d^\gamma$) from $A(Q_2)$ & ($-$0.118 , 0.120) & ($-$0.129 , 0.129) & 
($-$0.041 , 0.042) & ($-$0.273 , 0.273)\\ 
\hline 
{\it Im}($d^Z$) from $A(O_1)$ & ($-$0.234 , 0.236) & ($-$0.322 , 0.318) & 
($-$0.094 , 0.093) & ($-$3.35 , 3.24)\\  
\hline
{\it Im}($d^Z$) from $A(Q_2)$ & ($-$0.183 , 0.180) & ($-$0.283 , 0.282) & 
($-$2.03 , 1.99) & ($-$0.079 , 0.076)\\  
\hline 
\hline 
{\it Re}($d^\gamma$) from $A(O_2)$ & ($-$0.362 , 0.366) & ($-$0.311 , 0.308) & 
($-$0.111 , 0.111) & ($-$0.435 , 0.423)\\  
\hline
{\it Re}($d^\gamma$) from $A(Q_1)$ & ($-$0.170 , 0.168) & ($-$0.164 , 0.162) & 
($-$0.057 , 0.055) & ($-$0.281 , 0.277)\\  
\hline  
{\it Re}($d^Z$) from $A(O_2)$ & ($-$0.534 , 0.530) & ($-$0.611 , 0.612) & 
($-$1.71 , 1.72) & ($-$0.194 , 0.193)\\  
\hline
{\it Re}($d^Z$) from $A(Q_1)$ & ($-$0.223 , 0.218) & ($-$0.357 , 0.353) & 
($-$1.68 , 1.66) & ($-$0.098 , 0.093)
\end{tabular}
\caption{Expected 95\% CL bounds on $d^{Z,\gamma}$ from the asymmetry of
the observables at FLC for the dilepton decay mode.}
\label{table5}
\end{table}    

\begin{table}
\begin{tabular}{||c||c||c||c||c||}
Fraction ${\cal F}$ & 
${\cal P}^-_{e^-}$  & ${\cal P}^+_{e^-}$  & 
${\cal P}^-_{e^-} + {\cal P}^+_{e^-}$ & 
${\cal P}^-_{e^-} - {\cal P}^+_{e^-}$ \\ 
\hline
\hline
{\it Im}($d^\gamma$) from $O_1$ & ($-$0.053 , 0.053) & 
($-$0.075 , 0.074) & ($-$1.49 , 1.42) & ($-$0.022 , 0.021)\\
\hline
{\it Im}($d^\gamma$) from $\epsilon_2$ & ($-$0.056 , 0.058) & 
($-$0.067 , 0.071) & ($-$0.020 , 0.023) & ($-$0.239 , 0.257)\\
\hline 
{\it Im}($d^Z$) from $O_1$ & ($-$0.079 , 0.078) & 
($-$0.106 , 0.102) & ($-$0.032 , 0.029) & ($-$0.766 , 0.724)\\
\hline 
{\it Im}($d^Z$) from $\epsilon_2$ & ($-$0.095 , 0.088) & 
($-$0.176 , 0.169) & ($-$0.232 , 0.316) & ($-$0.046 , 0.039)\\
\hline 
\hline 
{\it Re}($d^\gamma$) from $O_2$ & ($-$0.126 , 0.125) & 
($-$0.107 , 0.107) & ($-$0.038 , 0.038) & ($-$0.149 , 0.150)\\
\hline 
{\it Re}($d^\gamma$) from $\epsilon_1$ & ($-$1.12 , 1.12) & 
($-$1.11 , 1.11) & ($-$0.375 , 0.375) & ($-$2.03 , 2.02)\\
\hline 
{\it Re}($d^Z$) from $O_2$ & ($-$0.189 , 0.179) & 
($-$0.248 , 0.243) & ($-$1.80 , 1.85) & ($-$0.076 , 0.069)\\
\hline
{\it Re}($d^Z$) from $\epsilon_1$ & ($-$1.52 , 1.52) & 
($-$2.76 , 2.73) & ($-$6.09 , 6.25) & ($-$0.702 , 0.687)
\end{tabular}
\caption{Expected 95\% CL bounds on $d^{Z,\gamma}$ from the expectation value 
(fraction ${\cal F}$) of the observables at FLC for the single lepton 
decay mode.}
\label{table6}
\end{table}    

\begin{table}
\begin{tabular}{||c||c||c||c||c||}
Asymmetry &
${\cal P}^-_{e^-}$  & ${\cal P}^+_{e^-}$  & 
${\cal P}^-_{e^-} + {\cal P}^+_{e^-}$ & 
${\cal P}^-_{e^-} - {\cal P}^+_{e^-}$ \\  
\hline
\hline
{\it Im}($d^\gamma$) from $A(O_1)$ & ($-$0.063 , 0.064) & 
($-$0.083 , 0.082) & ($-$0.487 , 0.469) & ($-$0.025 , 0.025)\\
\hline
{\it Im}($d^\gamma$) from $A(\epsilon_2)$ & ($-$0.055 , 0.056) & 
($-$0.061 , 0.064) & ($-$0.019 , 0.021) & ($-$0.158 , 0.172)\\
\hline 
{\it Im}($d^Z$) from $A(O_1)$ &  ($-$0.092 , 0.095) & 
($-$0.134 , 0.139) & ($-$0.039 , 0.038) & ($-$7.07 , 7.71)\\
\hline
{\it Im}($d^Z$) from $A(\epsilon_2)$ & ($-$0.101 , 0.094) & 
($-$0.158 , 0.152) & ($-$0.870 , 0.858) & ($-$0.045 , 0.039)\\
\hline 
\hline 
{\it Re}($d^\gamma$) from $A(O_2)$ &  ($-$0.153 , 0.156) & 
($-$0.128 , 0.127) & ($-$0.046 , 0.046) & ($-$0.172 , 0.166)\\ 
\hline
{\it Re}($d^\gamma$) from $A(\epsilon_1)$ & ($-$1.18  , 1.18) & 
($-$1.23 , 1.23) & ($-$0.406 , 0.406) & ($-$2.52 , 2.52)\\
\hline  
{\it Re}($d^Z$) from $A(O_2)$ & ($-$0.207 , 0.204) & 
($-$0.238 , 0.238) & ($-$0.680 , 0.694) & ($-$0.075 , 0.075)\\ 
\hline
{\it Re}($d^Z$) from $A(\epsilon_1)$ & ($-$1.57 , 1.57) & 
($-$9.78 , 9.93) & ($-$1.69 , 1.65) & ($-$1.03 , 1.06)
\end{tabular}
\caption{Expected 95\% CL bounds on $d^{Z,\gamma}$ from the asymmetry of
the observables at FLC for the single lepton decay mode.}
\label{table7}
\end{table}   

\end{document}